

\magnification=\magstep1
\baselineskip=1.8\baselineskip
\parskip=2pt plus 2pt

\newcount\RefNo \RefNo=1
\newbox\RefBox
\def\Ref [#1]{~[{\the\RefNo}]%
  \setbox\RefBox=\vbox{\unvbox\RefBox\leavevmode\item{[\the\RefNo]}
                       \frenchspacing #1}%
   \global\advance\RefNo by1 }
\def\Refc [#1]{~[{\the\RefNo,}%
  \setbox\RefBox=\vbox{\unvbox\RefBox\leavevmode\item{[\the\RefNo]}
                       \frenchspacing #1}%
   \global\advance\RefNo by1 }
\def\Refe [#1]{$\,${\the\RefNo}]%
  \setbox\RefBox=\vbox{\unvbox\RefBox\leavevmode\item{[\the\RefNo]}
                       \frenchspacing #1}%
   \global\advance\RefNo by1 }
\def\ShowReferences {\vfil\eject\leftline{\bf References}\smallskip
   \unvbox\RefBox}
\def\StoreRef #1{\edef#1{\the\RefNo}}

\def\Section #1\par{\bigbreak \leftline{\bf #1}\smallskip\noindent}%

\headline{\fiverm \hfil November 23, 1993}
\centerline{\bf Eternal inflation and the initial singularity}
\bigskip
\bigskip
\centerline{Arvind Borde$^{\ast\dag}$ and Alexander Vilenkin$^{\ddag}$}
\bigskip
\bigskip
\centerline{Institute of Cosmology}
\centerline{Department of Physics and Astronomy}
\centerline{Tufts University}
\centerline{Medford, MA 02155}

\vfootnote{$^{\ast}$}{{\baselineskip12pt
Permanent addresses: Long Island University, Southampton,
N.Y. 11968 and\hfil\break
\indent High Energy Theory Group, Brookhaven National Laboratory,
Upton, N.Y. 11973.\par}\vskip-\baselineskip}
\vfootnote{$^{\dag}$}{Electronic mail: borde@bnlcl6.bnl.gov}
\vfootnote{$^{\ddag}$}{Electronic mail: avilenki@pearl.tufts.edu}

\vfil
\centerline{\underbar{ABSTRACT}}
\smallskip It is shown that a physically reasonable spacetime that is
eternally inflating to the future must possess an initial singularity.

\vfil
\vfil

\leftline{PACS numbers: 04.50, 98.80.C}
\eject

\headline{\hfil}
\Section Introduction

Inflationary models of cosmology%
\Ref [For reviews see, for example,
A. D. Linde, {\it Particle Physics and Inflationary
Cosmology\/} (Harwood Academic, Chur Switzerland, 1990);
E. W. Kolb and M. S. Turner,
{\it The Early Universe\/} (Addison-Wesley, New York, 1990)]
yield a rather different picture of the full present universe
(as opposed to the part of it that we can directly observe) than that
given by the standard big bang cosmology. In current inflationary models
the universe consists of a number of isolated thermalized regions
embedded in a still-inflating background%
\Ref [The inflationary expansion is driven by the potential energy of a
scalar field $\varphi$, while the field slowly ``rolls down'' its
potential $V(\varphi)$. When $\varphi$ reaches the minimum of the
potential this vacuum energy thermalizes, and inflation is followed by
the usual radiation-dominated expansion. The evolution of the field
$\varphi$ is influenced by quantum fluctuations, and as a result
thermalization does not occur simultaneously in different parts of the
universe. Fluctuations in the thermalization time give rise to
small density fluctuations on observable scales, but result in
large deviations from homogeneity and isotropy on much larger scales.].
The boundaries of the thermalized regions expand into this background at
a speed approaching the speed of light, but the inflating domains that
separate them expand even faster and the thermalized regions never merge.
It is not possible to send a signal from the interior of a thermalized region
to the ``inflationary background'' in which it is embedded.

Theoretical work supported by computer simulations
suggests that this broad picture continues to describe the universe as
it evolves into the future%
\Refc [P. J. Steinhardt, in {\it The Very Early Universe}, edited by
G. W. Gibbons, S. W. Hawking and S. T. C. Siklos (Cambridge University
Press, Cambridge, England, 1983); A. Vilenkin, Phys. Rev. D {\bf 27},
2848 (1983); A. A. Starobinsky, in {\it Field Theory, Quantum Gravity
and Strings}, Proceedings of the Seminar series, Meudon and Paris, France,
1984--1985, edited by M. J. de Vega and N. Sanchez, Lecture Notes in
Physics Vol. 246 (Springer-Verlag, New York, 1986); A. D. Linde,
Phys. Lett. B {\bf 175}, 395 (1986).]%
\Refe [M. Aryal and A. Vilenkin, Phys.\ Lett.\ B {\bf 199}, 351 (1987);
A. S. Goncharov, A. D. Linde and V. F. Mukhanov, Int. J. Mod. Phys.
A {\bf 2}, 561 (1987); K. Nakao, Y. Nambu and M. Sasaki,
Prog. Theor. Phys. {\bf 80}, 1041 (1988); A. Linde, D. Linde and
A. Mezhlumian, Stanford preprint SU-ITP-93-13 (1993).].
Previously created regions expand
and new ones come into existence, but the universe does not fill up
entirely with thermalized regions. In other words, inflation is
{\it eternal to the future}.

A model in which the inflationary phase has no end and continuously
produces new islands of thermalization naturally leads to this question:
can this model also be extended to the infinite past, avoiding in this way
the problem of the initial singularity? The universe would then be
in a steady state of eternal inflation without a beginning.

The purpose of this paper is to show that this is in fact not possible in
future-eternal inflationary spacetimes as long as they obey some reasonable
physical conditions. Such models must necessarily possess initial
singularities; i.e., the inflationary universe must have had a
beginning.

A partial answer along these lines was previously given by Vilenkin%
\StoreRef{\Vilenkin}%
\Ref [A. Vilenkin, Phys.\ Rev.\ D {\bf 46}, 2355 (1992).] who
showed the necessity of a beginning in a two-dimensional spacetime
and gave a plausibility argument for four dimensions. The
broad question was also previously addressed by Borde\StoreRef{\Borde}%
\Ref [A. Borde, Cl.\ and Quant.\ Gravity {\bf 4}, 343 (1987).] who sketched
a general proof using the Penrose-Hawking-Geroch global techniques.
The proof given below will partly follow the sketch outlined in that paper.

\Section Statement of the result

{\sl A spacetime cannot be past null geodesically complete if it satisfies
the following conditions:
\item{A.} It is past causally simple.
\item{B.} It is open.
\item{C.} Einstein's equation holds, with a source that obeys the weak
  energy condition (i.e., the matter energy density is non-negative).
\item{D.} There is at least one point~$p$ such that for some
point~$q$ to the future of~$p$ the
volume of the difference of the pasts of~$q$ and~$p$ is finite.
\smallskip }
Observe that geodesic incompleteness is being taken as a signal that
there is a singularity. (A geodesic is incomplete if it cannot be continued
to arbitrarily large values of its affine parameter.)
This is the conventional approach in
singularity theorems.

It is worth noting that none
of the standard singularity theorems exactly fits the situation in
which we are interested: some of the theorems
assume the strong energy condition,
known to be violated in inflationary scenarios, and others place much
stronger restrictions on the global causal structure of spacetime than
we do here (through assumption~A). More significantly, assumption~D
is entirely new~-- as we shall see below, it captures a characteristic
aspect of future-eternal inflationary spacetimes.

\Section Analysis of the assumptions

Before we discuss the assumptions in detail, here is a summary:
Assumptions~A and~B are made solely for mathematical convenience.
The ultimate goal is to relax them (especially assumption~B).
Assumption~C holds in standard inflationary spacetimes, and is physically
quite reasonable. Assumption~D is necessary for
inflation to be future-eternal.

In our detailed discussion of the assumptions, and in the proof given below,
we will need to use some of the standard causal functions of the
`global techniques' approach to general relativity. This is
what we will need: A curve is called {\it causal\/} if it is everywhere
timelike or null (i.e., lightlike).
Let $p$ be a point in spacetime. The
{\it causal\/} and {\it chronological\/} pasts of $p$, denoted respectively
by J$^-(p)$ and I$^-(p)$, are defined as follows:

J$^-(p) = \{q:$ there is a future-directed causal curve
           from $q$ to $p\}$,

\noindent and

I$^-(p) = \{q:$ there is a future-directed timelike curve
           from $q$ to $p\}$.

The {\it past light cone\/} of~$p$ may then be defined as E$^-(p) =
{\rm J}^-(p) - {\rm I}^-(p)$. It may be shown\StoreRef{\HE}%
\Ref [For the detailed proofs of such standard global results, see, for
example, S. W. Hawking and G. F. R. Ellis, {\it The large scale structure of
space-time\/} (Cambridge University Press, Cambridge, England, 1973).]
that the boundaries of the two kinds of pasts of~$p$ are the same; i.e.,
$\dot {\rm J}^-(p) = \dot {\rm I}^-(p)$.
Further, it may be shown that E$^-(p) \subset \dot {\rm I}^-(p)$.
In general, however, E$^-(p) \ne \dot {\rm I}^-(p)$; i.e, the
past light cone of $p$ (as we have defined it here) is a subset of the
boundary of the past of $p$, but is not necessarily the full boundary
of this past.
This is illustrated in figure~1.

With this background in hand, we may now discuss the assumptions
in greater detail.

\smallskip\textindent{$\bullet$}
{\it Assumption A\/}: A spacetime is past causally simple
if E$^-(p) = \dot {\rm I}^-(p) \ne \emptyset$ for all points~$p$.
I.e., we exclude for now scenarios such as the one in figure~1.

\smallskip\textindent{$\bullet$}
{\it Assumption B\/}: A universe is open if it contains no
compact, achronal, edgeless hypersurfaces.
(A set is achronal if no two points in it can be connected by a
timelike curve.)

\smallskip\textindent{$\bullet$}
{\it Assumption C\/}: Einstein's equation is
$R_{ab}-{1\over 2}g_{ab}R = kT_{ab}$ (where $R_{ab}$ is the Ricci tensor
associated with the metric $g_{ab}$,
$R$ the scalar curvature, $T_{ab}$ the matter energy-momentum tensor, and
$k$ a positive constant in our conventions).
An observer with four-velocity $V^a$ will see a matter
energy density of $T_{ab}V^aV^b$. The weak energy condition is the
requirement that $T_{ab}V^aV^b \geq 0$ for all timelike vectors $V^a$.

This is a reasonable restriction on spacetime. (In fact, a much weaker
integral condition will do just as well for our purposes\Ref
[F. J. Tipler, J. Diff. Eq. {\bf 30}, 165 (1978); also see
{Ref.~[\Borde]} for an extension of these results.].)
It follows by continuity from the weak energy condition that
$T_{ab}N^aN^b \geq 0$ for all null vectors $N^a$ and from Einstein's
equation that $R_{ab}N^aN^b \geq 0$. It is this final form, sometimes
called the {\it null convergence condition}, that we will actually use.
(Thus our results will remain true even in other theories of gravity, as long
as the null convergence condition continues to hold.)

\smallskip\textindent{$\bullet$}
{\it Assumption D\/}: This condition was formulated~[\Vilenkin]
as a necessary condition for inflation to be future-eternal.
Consider a point $p$ that lies in the inflating
region; for inflation to be future-eternal there must be a non-zero
probability for there to be a point~$q$, at
a given timelike geodesic distance~$\delta$ to the
future of $p$, that also lies in the inflating region.
If thermalized regions expand at the speed of light, this means that there
should be no thermalization events
in the region ${\rm I}^-(q) - {\rm I}^-(p)$. Now,
the formation of thermalized regions is a stochastic process, and
it is reasonable to assume that there is a zero probability for
no such regions to form in an infinite spacetime volume.
This leads to assumption~D%
\Ref [Note that we are not assuming that there is
a non-zero probability
for a point $p$ to have a finite-volume past (i.e., a finite volume
for I$^-(p)$). Such an assumption would
mean that an arbitrarily chosen point has a finite probability to
be in the inflating region. It has been shown, however, that at least
in some models the inflating region is a fractal of dimension smaller than~4,
and therefore occupies a vanishing fraction of the total volume
(see {Ref.~[4]}).].
For a more detailed discussion of assumption~D see Ref.~[\Vilenkin].

\Section Proof

Suppose that a spacetime that obeys assumptions~A--D is past null
geodesically complete. We show in two steps that a contradiction ensues.

\textindent{1.} {\it A point $p$ that satisfies assumption~D has a finite
past light cone.} By a ``finite past light cone'' is meant a light cone
E$^-(p)$ such that every past-directed null geodesic that initially
lies in the cone leaves it a finite affine parameter distance to the
past of~$p$.

Suppose, to the contrary, that a null geodesic $\gamma$ lies in E$^-(p)$
an infinite affine parameter distance to the past of~$p$. Let~$v$ be
an affine parameter on~$\gamma$ chosen to increase to the past
and to have the value~0 at~$p$. Consider a small `conical' pencil
of null geodesics in E$^-(p)$ around~$\gamma$. Vary this pencil
an infinitesimal distance~$\Delta$ in the future time direction so that
the vertex is now at the point~$p'$. This sets up a congruence of
null geodesics; let $N^a$ be the tangent vector field to these geodesics.
(See figure~2.)

How do the null geodesics in this
congruence move away from or towards~$\gamma$? If one considers
a ``deviation vector'' $Z^a$ that connects points on~$\gamma$ to points on
nearby geodesics~[{\HE}], a straightforward calculation
yields the result that the only physically relevant variations
in~$Z^a$ come from its components in the spacelike 2-space
in E$^-(p)$ that is orthogonal~to~$N^a$%
\Ref [This may be seen by introducing a pseudo-%
orthonormal basis $\{N^a, L^a, X^a_1, X^a_2\}$ where $L^a$ is a null vector
such that $N^aL_a = 1$ (in a convention where the metric has
signature $(+, -, -, -)$) and $X^a_1$ and $X^a_2$
are unit spacelike vectors orthogonal to $N^a$ and~$L^a$.
The deviation vector $Z^a$ may be written
as $Z^a = n N^a + l L^a + x_1 X^a_1 + x_2 X^a_2$.
Now, it is possible to choose the affine parametrization so that $n=0$.
Further, $l=Z^aN_a$ and the derivative
of $l$ vanishes in the direction of $N^a$
(i.e., $N^bD_b(Z^aN_a)=0$).].

This means that the volume of the spacetime region occupied by the
geodesic congruence may be expressed as
$$ \delta V = \Delta \int_0^\infty {\cal A}(v)\, dv $$
where ${\cal A}(v)$ is the area of the spacelike cross section
of the light cone orthogonal to~$N^a$. The region whose volume we are
calculating is a subset of
$\overline{{\rm I}^-(q) - {\rm I}^-(p)}$ (i.e., the closure of
the difference of the pasts of~$q$ and~$p$)
and it must thus have a
finite volume. In order for this to happen, ${\cal A}$ must
decrease somewhere along~$\gamma$.

The propagation equation for $\cal A$ is~[{\HE}]
$$ \dot{\cal A} = \theta {\cal A} $$
where a dot represents a derivative with respect to~$v$ and $\theta =
D_aN^a$ is the divergence of the congruence. If ${\cal A}$ decreases
it follows that $\theta$ must become negative. But the propagation
equation for $\theta$ may be written as~[{\HE}]
$$ \dot\theta \leq -{1\over 2}\theta^2 - R_{ab}N^aN^b
              \leq -{1\over 2}\theta^2$$
(where we have used assumption~C in the last step).
If $\theta < 0 $ somewhere, it
follows that $\theta \to -\infty$ within a finite affine parameter distance.

The divergence of $\theta$ to $-\infty$ is a signal that the null
geodesics from~$p$ have refocused. It is a standard result in global
general relativity~[{\HE}] that points on such null geodesics beyond
the focal point enter the interior of the past light cone
(i.e., enter I$^-(p)$) and no longer lie in E$^-(p)$.

Thus the null cone E$^-(p)$ must be finite in the sense defined above.

\textindent{2.} {\it The result of step 1 contradicts assumption B.}
{}From causal simplicity it follows that E$^-(p)$ (being equal to the full
boundary of the past of $p$, $\dot{\rm I}^-(p)$) is an edgeless surface.
It is also achronal. (If two points on it can be connected by a timelike
curve, then the pastmost of the two points will lie inside I$^-(p)$, not
on its boundary~[{\HE}].) And step~1 implies that E$^-(p)$
is compact. These three statements taken together contradict assumption~B.

\Section Discussion

The conclusion to be drawn from this argument is that inflation does
not seem to avoid the problem of the initial singularity (although it
does move it back into an indefinite past). In fact, our analysis
of assumption~D suggests that almost all points in the inflating
region will have a singularity somewhere in their pasts.
The only way to deal with this
problem is probably to treat the universe quantum mechanically and
describe it by a wave function rather than by a classical spacetime.

The theorem that we have proved in this paper is based on several
assumptions which it would be desirable to further justify or relax.
The principal relaxation that is necessary is in assumption~B;
i.e., closed universes must also be accommodated. It may appear at first
sight that this will be difficult to achieve since assumption~B entered
into the proof above at a crucial place. However, all that we really
need is to exclude situations where null geodesics recross after running
around the whole universe (as they do in the static Einstein model).
If the reconvergence of the null cone discussed above
occurs on a scale smaller than the cosmological one, then essentially the
same argument goes through.  This approach to applying open universe
singularity theorems to closed universes has been outlined previously
by Penrose\Ref [R. Penrose, in {\it Battelle Rencontres}, edited by
C. M. DeWitt and J. A. Wheeler (W. A. Bejamin, New York, 1968).]
and it will be discussed in detail separately as will the relaxation
of assumption~A.

Assumption~D, which plays a central role in our argument, requires
further justification. It rests on the assumptions that (i)~the boundaries
of thermalized regions expand at speeds approaching the speed of light,
and (ii)~that the probability of finding no thermalized regions in an
infinite spacetime volume vanishes. Both assumptions are plausible
and are known to be true in the original inflationary scenario%
\Refc [A. H. Guth, Phys. Rev. D {\bf 23}, 347 (1981).]%
\Refe [K. Sato, Mon. Not. R. Astr. Soc., {\bf 195}, 467 (1981).]
in which the role of ``thermalization'' is played by bubble nucleation.
It would be interesting, however, to determine the exact conditions of
validity for these assumptions and to investigate the possibility
of relaxing them.

\ShowReferences
\vfil\eject

\leftline{\bf Figures}
\bigskip
\noindent {\bf Figure 1:} An example of a spacetime that is not
causally simple. The two thick horizontal lines are identified, allowing the
point $q$ to send a signal to $p$ as shown (along the curve~$\gamma$).
The boundary of the past of $p$
then consists of more than just the past light cone of~$p$; i.e.,
$\dot{\rm I}^-(p) - {\rm E}^-(p)$ is not empty.
\bigskip
\noindent {\bf Figure 2:} The volume of interest.

\vfil\eject
\bye